\documentclass[prd,nofootinbib,preprint,superscriptaddress,twocolumn,10pt]{revtex4}
\pdfoutput=1
\usepackage[T1]{fontenc}
\usepackage{amsmath,amssymb}
\usepackage{braket}
\usepackage{epsfig}
\usepackage{graphicx}
\usepackage[usenames,dvipsnames]{color}
\usepackage{subfigure}
\usepackage{slashed}
\usepackage[colorlinks,citecolor=blue]{hyperref}
\usepackage{pdfpages}
\usepackage{color}
\usepackage{comment}
\begin{document}


\title{Electromagnetic Dirac Cogenesis}

\author{Debasish Borah}
\email{dborah@iitg.ac.in}
\affiliation{Department of Physics, Indian Institute of Technology Guwahati, Assam 781039, India}
\affiliation{Pittsburgh Particle Physics, Astrophysics, and Cosmology Center, Department of Physics and Astronomy, University of Pittsburgh, Pittsburgh, PA 15260, USA}

\author{Arnab Dasgupta}
\email{arnabdasgupta@pitt.edu}
\affiliation{Pittsburgh Particle Physics, Astrophysics, and Cosmology Center, Department of Physics and Astronomy, University of Pittsburgh, Pittsburgh, PA 15260, USA}

\begin{abstract}
We propose a novel cogenesis mechanism by utilising the two-body decay of heavy vector-like fermions to dark matter (DM) $\chi$ and right chiral part of light Dirac neutrino $\nu_R$ via the electromagnetic dipole operator. This leads to generation of asymmetry in dark fermion $\chi$ as well as $\nu_R$  with the latter getting transferred to left-handed lepton doublets via Yukawa interactions with a neutrinophilic Higgs doublet. While lepton asymmetry is converted into baryon asymmetry of the Universe via electroweak sphalerons, the dark fermion asymmetry results in asymmetric dark matter. Since CP asymmetries in lepton and dark sector are equal and opposite due to net lepton number conservation, DM mass is restricted to a fixed value $\sim \mathcal{O}(1)$ GeV. Long-lived nature of DM keeps indirect detection prospects at gamma-ray telescopes alive while thermalised light Dirac neutrinos lead to observable dark radiation at cosmic microwave background (CMB) experiments. Heavy vector-like fermions can be probed at terrestrial experiments via their electromagnetic dipole interactions.
\end{abstract}
\maketitle
\section{Introduction}
The matter content in the present Universe is dominated by dark matter (DM) leaving only $\sim 20\%$ of the matter density to be composed of ordinary or visible matter \cite{Planck:2018vyg, ParticleDataGroup:2024cfk}. While the origin of DM remains a puzzle, the asymmetric nature of the visible or baryonic matter leads to another mystery, referred to as the baryon asymmetry of the Universe (BAU). While the standard model (SM) of particle physics does not provide any solution to the puzzles of DM and BAU, several beyond standard model (BSM) proposals have been studied in the literature to accommodate these observations. Among them, the weakly interacting massive particle (WIMP) \cite{Kolb:1990vq, Jungman:1995df, Bertone:2004pz} paradigm  for DM and baryogenesis \cite{Weinberg:1979bt, Kolb:1979qa} as well as leptogenesis \cite{Fukugita:1986hr} for BAU have been the most widely studied ones. On the other hand, the similar order of magnitude abundances of DM and BAU namely, $\Omega_{\rm DM} \approx 5\,\Omega_{\rm B}$ has also led to several works finding a common origin or cogenesis mechanism. Some widely studied cogenesis mechanisms include, but not limited to, asymmetric dark matter (ADM) \cite{Nussinov:1985xr, Davoudiasl:2012uw, Petraki:2013wwa, Zurek:2013wia, Barman:2021ost, Borah:2024wos}, baryogenesis from DM annihilation \cite{Yoshimura:1978ex, Barr:1979wb, Baldes:2014gca, Cui:2011ab, Bernal:2012gv, Bernal:2013bga, Kumar:2013uca, Racker:2014uga, Borah:2018uci, Borah:2019epq, Dasgupta:2019lha, Mahanta:2022gsi}, Affleck-Dine cogenesis \cite{Roszkowski:2006kw, Seto:2007ym, Cheung:2011if, vonHarling:2012yn, Borah:2022qln, Borah:2023qag}. Recently, there have also been attempts to generate DM and BAU together via a first-order phase transition (FOPT) by utilising the mass-gain mechanism \cite{Baldes:2021vyz, Azatov:2021irb, Borah:2022cdx, Borah:2023saq}, forbidden decay of DM \cite{Borah:2023god} or bubble filtering \cite{Borah:2025wzl}. 

In this letter, we propose a novel cogenesis mechanism where equal and opposite CP asymmetries are generated in light Dirac neutrinos and a Dirac fermion dark fermion $\chi$ from out-of-equilibrium decay of heavy vector-like fermions $N$. Unlike typical seesaw models for neutrino masses, we utilise electromagnetic dipole operators involving $N, \chi$ and right chiral part of Dirac neutrino $\nu_R$ to facilitate the decay of $N$ thereby generating the required asymmetries. The asymmetry generated in $\nu_R$ is then transferred to left-handed lepton doublets via Yukawa interactions with a neutrinophilic Higgs doublet \cite{Heeck:2013vha, Borah:2022qln}. The lepton asymmetry generated before the sphaleron decoupling epoch gets converted into baryon asymmetry via electroweak sphalerons. The asymmetry generated in $\chi$ leads to a long-lived asymmetric dark matter. Unlike typical electromagnetic leptogenesis \cite{Bell:2008fm, Choudhury:2011gbi} requiring three-body decay, here we utilise two-body decay \cite{Borah:2025oqj} which is possible due to the gauge singlet nature of the particles involved. Due to the absence of any net lepton number violation, the lepton asymmetry is related to the dark fermion asymmetry thereby restricting DM mass to a fixed value, in the spirit of the original asymmetric dark matter formalism. We constrain the scale of cogenesis and strength of the dipole operator from the requirements of generating the desired BAU and DM relic. The same parameter space gets also constrained from the lifetime criteria of DM which can decay into two photons and a neutrino. While long-lived nature of DM keeps indirect detection prospects alive, heavy vector-like fermions can be probed at terrestrial experiments via their dipole portal interactions.

This paper is organised as follows. In section \ref{sec1}, we outline our model followed by the details of cogenesis in section \ref{sec1a}. In section \ref{sec2}, we discuss our numerical results related to cogenesis. We briefly comment on the detection prospects of the model in section \ref{sec3} and finally conclude in section \ref{sec4}.

\section{The framework}
\label{sec1}
We consider a scenario where the standard model fermion content is extended by two types of vector-like singlet neutral fermions $N_{L,R}, \chi_{L,R}$ and a singlet chiral fermion $\nu_R$. While $\nu_R$ combines with the left chiral neutrinos $\nu_L$ to form Dirac neutrinos at tree level via a neutrinophilic Higgs doublet $H_2$, the heavy vector-like fermion $N$ has an effective dipole coupling with $\nu_R$ as well as $\chi$. A global $U(1)_D$ symmetry, similar to lepton number, is assumed to be conserved which prevents any Majorana terms. Given this, the relevant Lagrangian invariant under the SM gauge symmetry and $U(1)_D$ is given by 
\begin{align}
   -\mathcal{L} & \supset \frac{\lambda_{i\alpha}}{\Lambda} \overline{N_{iL}} \sigma_{\mu \nu} \nu_{R\alpha} B^{\mu \nu}+\frac{h_{Li}}{\Lambda} \overline{N_{iR}} \sigma_{\mu \nu} \chi_{L} B^{\mu \nu}  \nonumber \\
   & +\frac{h_{Ri}}{\Lambda} \overline{N_{iL}} \sigma_{\mu \nu} \chi_{R} B^{\mu \nu}+ y_{\alpha \beta} \overline{L}_\alpha \tilde{H_2} \nu_{R_\beta} + \nonumber \\
   & + M_i \overline{N_{iL}} N_{iR} + m_\chi \overline{\chi_L} \chi_R+ {\rm h.c.} 
   \label{eq1}
\end{align}
where $B^{\mu \nu} = \partial^\mu B^\nu-\partial^\nu B^\mu$ is the field strength tensor of $U(1)_Y$ in the SM. Neutrinos acquire sub-eV Dirac mass $M_D = y v_2/\sqrt{2}$ after the second Higgs doublet $H_2$ acquires a tiny vacuum expectation value (VEV) $v_2$ \cite{Davidson:2010sf}.

\begin{figure}
    \includegraphics[scale=0.5]{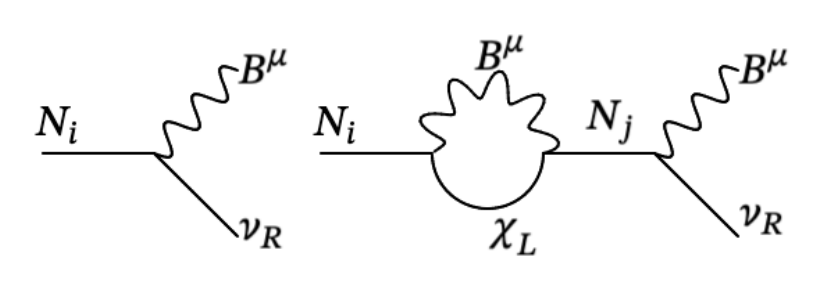}
    \caption{Feynman diagrams of processes responsible for generating asymmetry in $\nu_R$.}
    \label{fig1}
\end{figure}

\begin{figure}
    \includegraphics[scale=0.5]{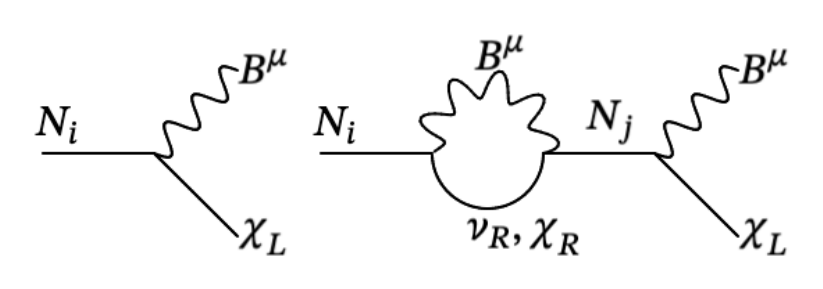}
        \includegraphics[scale=0.5]{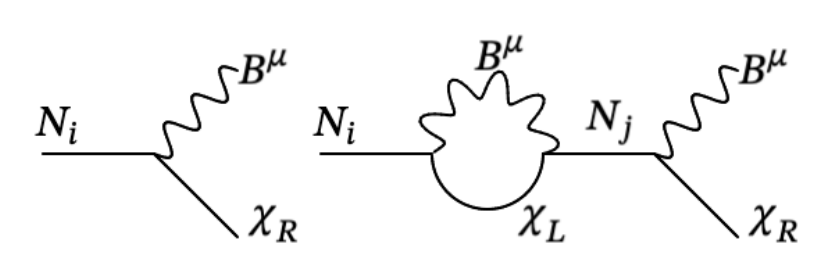}
    \caption{Feynman diagrams of processes responsible for generating asymmetry in $\chi$.}
    \label{fig2}
\end{figure}

The heavy vector-like fermion can decay out-of-equilibrium to create asymmetries in $\nu_R, \chi$ as shown in Fig. \ref{fig1} and Fig. \ref{fig2} respectively. This is similar to the idea of Dirac leptogenesis \cite{Dick:1999je, Murayama:2002je, Cerdeno:2006ha, Borah:2016zbd, Barman:2022yos, Barman:2023fad, Borboruah:2024lli} where equal and opposite CP asymmetries are created in left- and right-handed neutrino sectors. Recently, a similar cogenesis mechanism with equal and opposite CP asymmetries for lepton and dark matter was proposed \cite{Bandyopadhyay:2025hoc} with effective operators similar to axion-like particles (ALP) instead of the electromagnetic dipole considered here. Denoting the CP asymmetries as $\epsilon_{\nu_R}, \epsilon_{\chi_L}, \epsilon_{\chi_R}$ and given the fact that there is no net lepton number violation due to pure Dirac nature of all the fermions, we have 
\begin{equation}
    \epsilon_{\nu_R}+\epsilon_{\chi_L}+\epsilon_{\chi_R}=0.
    \label{eq:sum1}
\end{equation}
The CP asymmetry $\epsilon_x$ is defined as
\begin{align}
    \epsilon_x = \frac{\Gamma (N_i \rightarrow x B^\mu) - \Gamma (N_i \rightarrow \overline{x} B^\mu)}{\Gamma_{i}}
\end{align}
where the partial decay widths at tree level are given by 
\begin{equation}
    \Gamma_0 (N_i \rightarrow \nu_R B^\mu) =\Gamma_0 (N_i \rightarrow \overline{\nu_R} B^\mu)= \frac{(\lambda^\dagger \lambda)_{ii}}{4\pi} \frac{M^3_i}{\Lambda^2},
    \label{decaywidth}
\end{equation}
\begin{equation}
    \Gamma_0 (N_i \rightarrow \chi_R B^\mu) =\Gamma_0 (N_i \rightarrow \overline{\chi_R} B^\mu)= \frac{(h^\dagger_R h_R)_{ii}}{4\pi} \frac{M^3_i}{\Lambda^2},
    \label{decaywidth}
\end{equation}
\begin{equation}
    \Gamma_0 (N_i \rightarrow \chi_L B^\mu) =\Gamma_0 (N_i \rightarrow \overline{\chi_L} B^\mu)= \frac{(h^\dagger_L h_L)_{ii}}{4\pi} \frac{M^3_i}{\Lambda^2},
    \label{decaywidth}
\end{equation}
where we have assumed the final state particles to be massless. The CP asymmetries at one loop, corresponding to the processes shown in Fig. \ref{fig1} and Fig. \ref{fig2} are given by
\begin{align}
    \epsilon_{\nu_R} &= \frac{{\rm Im}[\lambda^*_{j\alpha}h_{Lj}h^*_{Li}\lambda_{i\alpha}]}{|h_{Li}|^2\widetilde{\Gamma}_{\rm tot}}\Gamma_0(N_i \rightarrow \chi_L B^\mu)\frac{r_{ji}}{1-r^2_{ji}} \nonumber \\
    &+ \frac{{\rm Im}[\lambda^*_{j\alpha}h_{Rj}h^*_{Ri}\lambda_{i\alpha}]}{|h_{Li}|^2\widetilde{\Gamma}_{\rm tot}}\Gamma_0(N_i \rightarrow \chi_R B^\mu)\frac{1}{1-r^2_{ji}},
\end{align}
\begin{align}
    \epsilon_{\chi_R} &= \frac{{\rm Im}[h^*_{Rj} h_{Lj} h^*_{Li}h_{Ri}]}{|h_{Li}|^2\widetilde{\Gamma}_{\rm tot}}\Gamma_0(N_i \rightarrow \chi_L B^\mu)\frac{r_{ji}}{1-r^2_{ji}} \nonumber \\
    &+ \frac{{\rm Im}[h^*_{Rj} \lambda_{j\alpha} \lambda^*_{i\alpha} h_{Ri}]}{|\lambda_{i\alpha}|^2 \widetilde{\Gamma}_{\rm tot}}\Gamma_0(N_i \rightarrow \nu_R B^\mu)\frac{1}{1-r^2_{ji}},
\end{align}
\begin{align}
    \epsilon_{\chi_L} &= \frac{{\rm Im}[h^*_{Lj} h_{Rj} h^*_{Ri}  h_{Li}]}{|h_{Ri}|^2\widetilde{\Gamma}_{\rm tot}}\Gamma_0(N_i \rightarrow \chi_R B^\mu)\frac{r_{ji}}{1-r^2_{ji}} \nonumber \\
    &+ \frac{{\rm Im}[h^*_{Lj} \lambda_{j\alpha} \lambda^*_{i\alpha} h_{Li}]}{|\lambda_{i\alpha}|^2\widetilde{\Gamma}_{\rm tot}}\Gamma_0(N_i \rightarrow \nu_R B^\mu)\frac{r_{ji}}{1-r^2_{ji}},
\end{align}
where $r_{ji} = M^2_j/M^2_i$ and $\widetilde{\Gamma}_{\rm tot} = |h_{Ri}|^2 + |h_{Li}|^2 + |\lambda_{i\alpha}|^2$. Using the above expressions for CP asymmetries, it is clear that $\epsilon_{\nu_R}+\epsilon_{\chi_L}+\epsilon_{\chi_R}=0$ from the requirement of total lepton number conservation.

\section{Cogenesis  baryon and dark matter}
\label{sec1a}
The Boltzmann equations can be written as 
\begin{align}
   \dfrac{d \eta_{\Delta \nu_R}}{d z} &= \frac{1}{n_\gamma \mathcal{H}z} \bigg [ \epsilon_{\nu_R} \left ( \frac{\eta_{N_{1R}}}{\eta^{\rm eq}_{N_{1R}}}-1 \right ) \gamma (N_{1R} \rightarrow B^\mu \nu_R) \nonumber \\
   & -\frac{\eta_{\Delta \nu_R}}{2\eta^{\rm eq}_{\nu_R}} \gamma (N_{1R} \rightarrow B^\mu \nu_R) \nonumber \\
   & + \left ( \frac{\eta_{\Delta \chi_R}}{\eta^{\rm eq}_{\chi_R}}-\frac{\eta_{\Delta \nu_R}}{\eta^{\rm eq}_{\nu_R}} \right) \gamma (\chi_R B^\mu \rightarrow \nu_R B^\mu) \nonumber \\
   & + \left ( \frac{\eta_{\Delta \chi_L}}{\eta^{\rm eq}_{\chi_L}}-\frac{\eta_{\Delta \nu_R}}{\eta^{\rm eq}_{\nu_R}} \right) \gamma (\chi_L B^\mu \rightarrow \nu_R B^\mu) \bigg ],
\end{align}
\begin{align}
   \dfrac{d \eta_{\Delta \chi_R}}{d z} &= \frac{1}{n_\gamma \mathcal{H}z} \bigg [ \epsilon_{\chi_R} \left ( \frac{\eta_{N_{1R}}}{\eta^{\rm eq}_{N_{1R}}}-1 \right ) \gamma (N_{1R} \rightarrow B^\mu \chi_R) \nonumber \\
   & -\frac{\eta_{\Delta \chi_R}}{2\eta^{\rm eq}_{\chi_R}} \gamma (N_{1R} \rightarrow B^\mu \chi_R) \nonumber \\
   & + \left ( \frac{\eta_{\Delta \nu_R}}{\eta^{\rm eq}_{\nu_R}}-\frac{\eta_{\Delta \chi_R}}{\eta^{\rm eq}_{\chi_R}} \right) \gamma (\nu_R B^\mu \rightarrow \chi_R B^\mu) \nonumber \\
   & + \left ( \frac{\eta_{\Delta \chi_L}}{\eta^{\rm eq}_{\chi_L}}-\frac{\eta_{\Delta \chi_R}}{\eta^{\rm eq}_{\chi_R}} \right) \gamma (\chi_L B^\mu \rightarrow \chi_R B^\mu) \bigg ],
\end{align}
\begin{align}
   \dfrac{d \eta_{\Delta \chi_L}}{d z} &= \frac{1}{n_\gamma \mathcal{H}z} \bigg [ \epsilon_{\chi_L} \left ( \frac{\eta_{N_{1L}}}{\eta^{\rm eq}_{N_{1L}}}-1 \right ) \gamma (N_{1L} \rightarrow B^\mu \chi_L) \nonumber \\
   & -\frac{\eta_{\Delta \chi_L}}{2\eta^{\rm eq}_{\chi_L}} \gamma (N_{1L} \rightarrow B^\mu \chi_L) \nonumber \\
   & + \left ( \frac{\eta_{\Delta \nu_R}}{\eta^{\rm eq}_{\nu_R}}-\frac{\eta_{\Delta \chi_L}}{\eta^{\rm eq}_{\chi_L}} \right) \gamma (\nu_R B^\mu \rightarrow \chi_L B^\mu) \nonumber \\
   & + \left ( \frac{\eta_{\Delta \chi_R}}{\eta^{\rm eq}_{\chi_R}}-\frac{\eta_{\Delta \chi_L}}{\eta^{\rm eq}_{\chi_L}} \right) \gamma (\chi_R B^\mu \rightarrow \chi_L B^\mu) \bigg ],
\end{align}
\begin{align}
    \dfrac{d \eta_{N_{1L}}}{d z} &= -\frac{1}{n_\gamma \mathcal{H}z} \bigg [ \left ( \frac{\eta_{N_{1L}}}{\eta^{\rm eq}_{N_{1L}}} -1 \right) \gamma (N_{1L} \rightarrow B^\mu \chi_L) \bigg ],
\end{align}
\begin{align}
    \dfrac{d \eta_{N_{1R}}}{d z} &= -\frac{1}{n_\gamma \mathcal{H}z} \bigg [ \left ( \frac{\eta_{N_{1R}}}{\eta^{\rm eq}_{N_{1R}}} -1 \right) \gamma (N_{1R} \rightarrow B^\mu \chi_R) \nonumber \\
    & + \left ( \frac{\eta_{N_{1R}}}{\eta^{\rm eq}_{N_{1R}}} -1 \right) \gamma (N_{1R} \rightarrow B^\mu \nu_R) \bigg ],
\end{align}
where $\eta_x=n_x/n_\gamma$ is the comoving number density of $x$ and $z=M_1/T$. In the above equations, the thermally-averaged reaction densities for the $2 \rightarrow 2$ scattering processes are defined as 
\begin{eqnarray}
    \gamma_{ij\longrightarrow kl } = \frac{ T}{64 \pi^{4}} \int_{(m_{i}+m_{j})^{2}}^{\infty} ds \sqrt{s}K_{1} \left( \sqrt{s}/T \right) \hat{\sigma}(s),
\end{eqnarray}
where $\hat{\sigma}(s)$ is the reduced cross section for the process and is given by 
\begin{equation}
\hat{\sigma} (s)= 8 \left[ (p_{i}.p_{j})^{2}-m_{i}^{2}m_{j}^{2} \right] \sigma (s).
\end{equation}
The $\gamma_{ij \longrightarrow kl}$ is related to the $\langle  \sigma v \rangle_{ij\longrightarrow kl}$ can be identified to be 
\begin{equation}
\gamma_{ij \longleftarrow kl}=n_{i}^{\rm eq}n_{j}^{\rm eq}\langle  \sigma v \rangle_{ij \longrightarrow kl}
\end{equation}
where $n_{i}^{\rm eq}$ and $n_{j}^{\rm eq}$ are the equilibrium number densities of $i$ and $j$ respectively.  For decay $A\longrightarrow B$ the reaction densities are defined as 
\begin{equation}
    \gamma_{A\longrightarrow B}=n_{A}^{\rm eq} \frac{K_{1}\left( m_{A}/T \right)}{K_{2} \left(  m_{B}/T \right)}\Gamma_{A\longrightarrow B}.
\end{equation}
In the above expressions, $K_i$'s denote modified Bessel functions of the second kind.

In order to prevent the asymmetry in $\nu_R$ from being washed out, one has to ensure that the processes $\nu_R B^\mu \leftrightarrow \chi B^\mu$ do not equilibrate. This can be taken care of by keeping the rates of these processes below the Hubble expansion rate of the Universe $\mathcal{H}$. The corresponding out-of-equilibrium condition can be stated as
\begin{equation}
    \Gamma^{\rm wo}_{\rm LR} \equiv \frac{\lvert \lambda^\dagger_{1\alpha} h_{L1} \rvert^2 M^2_1}{\Lambda^4} T^3 < \mathcal{H}= \sqrt{\frac{8\pi^3g_*}{90}}\frac{T^2}{M_{\rm Pl}}.
\end{equation}

While we ensure that $\nu_R$ asymmetry does not get equilibrated with $\chi$ asymmetry, it is crucial to ensure that the former gets transferred to the lepton doublets. As mentioned earlier, this is made possible via the Dirac Yukawa coupling of neutrinos with the neutrinophilic Higgs $H_2$. Demanding this Yukawa interaction to be in equilibrium prior to the sphaleron decoupling epoch leads to a lower bound 
\begin{equation}
    y \gtrsim 10^{-8}
\end{equation}
ensuring the transfer of $\nu_R$ asymmetry into lepton doublets. Once the asymmetry in $\nu_R$ is transferred to the left-handed lepton doublets, the electroweak sphaleron processes convert the $B-L$ asymmetry into baryon asymmetry as
\begin{align}
   \eta_{\rm B}= \frac{8 N_f + 4 N_\textbf{H}}{22 N_f + 13 N_\textbf{H}}\frac{\eta_{\rm B-L}}{S} = \frac{C_{\rm sph}}{S} \eta_{\rm B-L}\,,\label{eqn:sphconv}   
\end{align}
where $C_{\rm sph}= \frac{8}{23}$ for two Higgs doublets $N_\textbf{H}=2$ and three fermion generations $N_f=3$. The factor $S$ accounts for the change in the relativistic degrees of freedom from the scale of leptogenesis until recombination and comes out to be $S=\frac{106.75}{3.91}\simeq 27.3$.

\begin{figure}
    \includegraphics[scale=0.5]{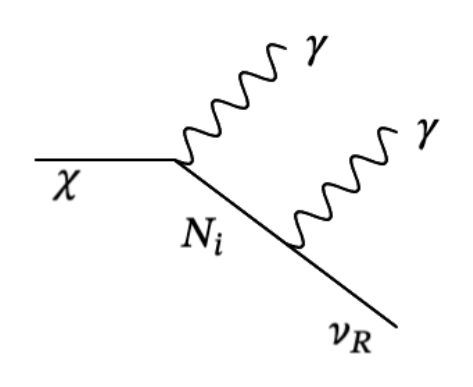}
    \caption{Three-body decay of dark matter $\chi$.}
    \label{fig4}
\end{figure}

On the other hand, if $\chi$ has to satisfy the observed dark matter relic, it needs to be long-lived on cosmological scales. However, $\chi$ can have three-body decay into $\nu_R \gamma \gamma$ mediated by heavy vector-like fermion $N_i$, as shown in Fig. \ref{fig4}. The corresponding decay width can be found as 
\begin{equation}
    \Gamma (\chi \rightarrow \gamma \gamma \nu_R) = \frac{37 \lambda^2_{i\alpha} h^2_{Li}}{2560\pi^3} \frac{m^7_\chi}{\Lambda^4 M^2_i}.
\end{equation}
The dark fermion $\chi$ also acquires a one-loop mass given by
\begin{equation}
    m^{\rm loop}_\chi \approx \frac{1}{16\pi^2} \frac{1}{\Lambda^2} h_{Li} M^3_{i} h_{Ri}. 
\end{equation}
Similar one-loop contribution also arises for off-diagonal $\chi-\nu_R$ terms. However, in the limit of sub-eV Dirac neutrino mass $M_D \ll m_\chi, m^{\rm loop}_{\chi \nu}$, it does not lead to additional constraints from neutrino data or $\chi$ lifetime requirements.

\begin{figure}
    \includegraphics[scale=0.5]{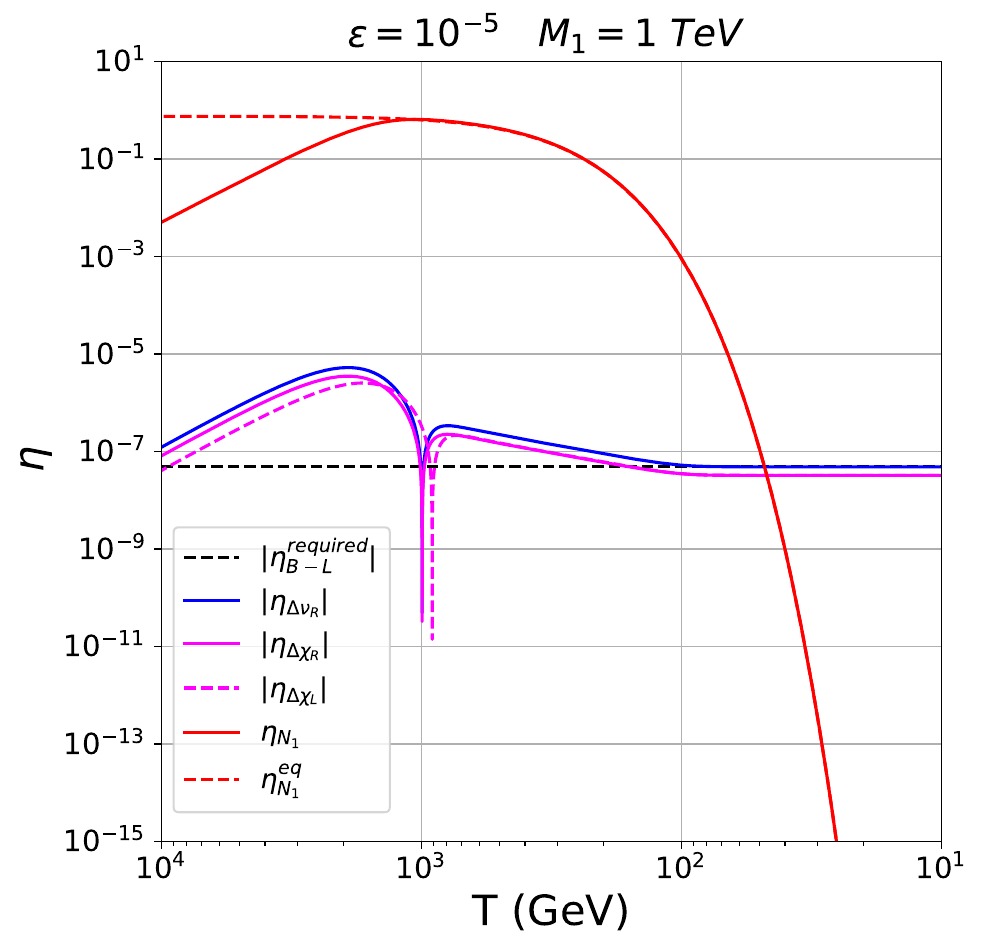}
    \caption{Evolution of comoving number densities for $M_1=1$ TeV, $\Lambda=1.5 \times 10^9$ GeV, $\lambda_{i\alpha}, h_{Li}, h_{Ri} \sim 1$.}
    \label{fig5}
\end{figure}

\section{Results and Discussion}
\label{sec2}
Fig. \ref{fig5} shows the evolution of the absolute values of the comoving asymmetries in $\nu_R, \chi_R, \chi_L$ for fixed values of CP asymmetry parameters, heavy fermion mass and axion decay constant. The same figure also shows the comoving number density of $N_1$ with a comparison to its equilibrium value. When $\eta_{N_1}$ reaches its equilibrium value $\eta^{\rm eq}_{N_1}$, the asymmetries acquire a sign flip leading to the dip in their evolution. Subsequently, washouts due to inverse decay reduces the asymmetries further before saturating to their asymptotic values. For this particular benchmark, we choose $\epsilon_{\nu_R}=3 \epsilon, \epsilon_{\chi_R} = -2\epsilon, \epsilon_{\chi_L}=-\epsilon$ with $\epsilon=10^{-5}$. While the net CP asymmetry is zero by unitarity, the asymmetry generated in $\nu_R$ before the sphaleron decoupling epoch $T_{\rm sph} \sim 131.7$ GeV \cite{DOnofrio:2014rug} is consistent with the required lepton asymmetry to generate the observed BAU. The dashed horizontal line in Fig. \ref{fig5} corresponds to the required lepton asymmetry. The net dark asymmetry adds up to generate an asymmetric $\chi$ abundance at present epoch $(T=T_0)$ given by 
\begin{equation}
    \Omega_{\chi} (T_0) = m_\chi \frac{(\eta_{\Delta \chi_L}+\eta_{\Delta \chi_R})}{S} \frac{n_\gamma (T_0)}{\rho_{\rm cr}(T_0)}
\end{equation}
where $S$ is the same dilution factor used in Eq. \eqref{eqn:sphconv} and $\rho_{\rm cr}(T_0)=\frac{3\mathcal{H}^2_0}{8\pi G}$ is the critical density of the Universe at present. Since $ \lvert \eta_{\Delta \chi_L}+\eta_{\Delta \chi_R} \rvert \sim \eta_{\Delta \nu_R}$ and $\Omega_{\rm DM} \approx 5.36 \Omega_B$ \cite{Planck:2018vyg}, we get $m_\chi \leq 1.9 m_p$ such that $\Omega_\chi \leq \Omega_{\rm DM}$. 

\begin{figure*}
    \centering
    \includegraphics[width=0.49\linewidth]{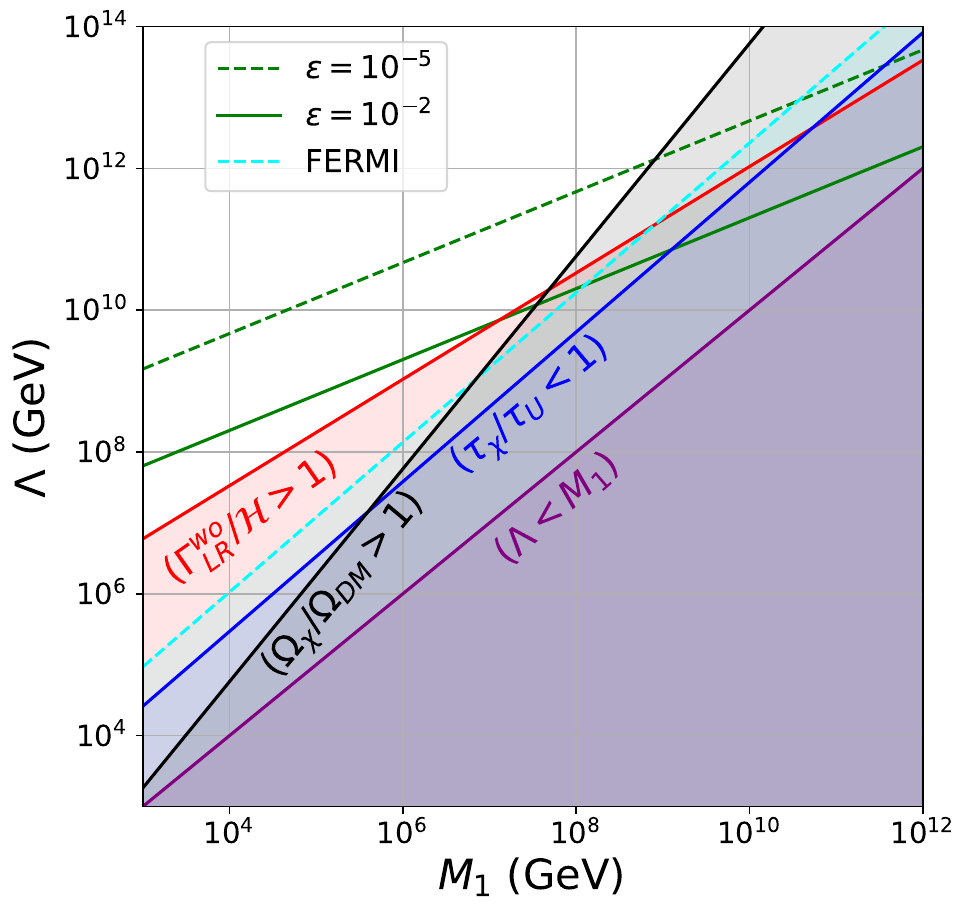}
    \includegraphics[width=0.49\linewidth]{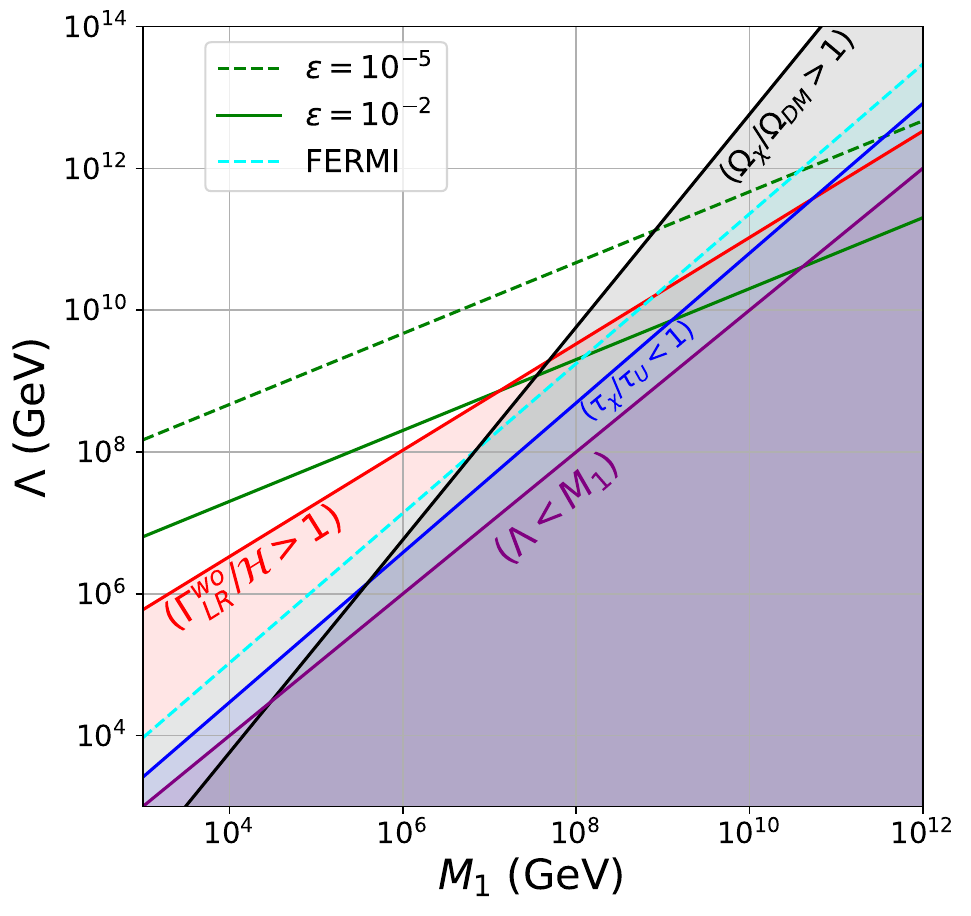}
    \caption{Allowed parameter space of electromagnetic Dirac cogenesis in $\Lambda-M_1$ plane considering dimensionless couplings $\lambda_\alpha, h_{L,R} $ to be $1$ (left panel) and $0.1$ (right panel). The shaded regions are disfavored due to different constraints, leaving only the white region allowed for successful cogenesis.}
    \label{fig:sum}
\end{figure*}

Fig. \ref{fig:sum} shows the summary of parameter space in $\Lambda-M_1$ plane. The green colored contours correspond to the parameter space consistent with the observed baryon asymmetry for two different values of CP asymmetry parameter. For a fixed $M_1$, smaller values of $\Lambda$ correspond to larger washout from inverse decay requiring larger CP asymmetry to generate the desired asymmetry. The red shaded region labeled as $\Gamma^{\rm wo}_{\rm LR}/\mathcal{H}>1$ is disfavored due to strong washout from equilibration of lepton and dark fermion asymmetries. The grey shaded region is ruled out due to overproduction of dark matter $\Omega_\chi/\Omega_{\rm DM}>1$. The blue shaded region is disfavored as DM lifetime is less than the age of the Universe. A stronger bound on lifetime arises from non-observation of diffuse gamma-rays by indirect detection experiments like FERMI \cite{Fermi-LAT:2012edv}, as analysed for three-body decays of DM with two-photon final states in \cite{Essig:2013goa}. This leads to a lower bound on $\Lambda$, shown by the cyan colored contour. The purple shaded region labeled as $\Lambda < M_1$ corresponds to the parameter space where EFT description is invalid. While the left panel corresponds to dimensionless couplings like $\lambda_\alpha, h_{L, R}$ to be unity, the right panel considers them to be $0.1$. This leads to more allowed parameter space on the right panel compared to the left panel of Fig. \ref{fig:sum}.

\section{Detection Aspects}
\label{sec3}
One of the unique predictions of our cogenesis mechanism is the mass of DM $m_\chi \sim 1.9 m_p$ with a characteristic decay channel $\chi \rightarrow \gamma \gamma \nu_R$. While such continuum gamma-ray emission from three-body decay of GeV scale DM can be constrained from FERMI \cite{Fermi-LAT:2012edv} and EGRET \cite{Hunter:1997qec} observations, future low-energy gamma-ray telescopes like GECCO \cite{Orlando:2021get}, e-ASTROGAM \cite{e-ASTROGAM:2017pxr}, AMEGO \cite{Kierans:2020otl}, MAST \cite{Dzhatdoev:2019kay}, AdEPT \cite{Hunter:2013wla}, PANGU \cite{Wu:2014tya}, GRAMS \cite{Aramaki:2019bpi} will be able to probe a the currently allowed parameter space even further, as summarised recently in \cite{ODonnell:2024aaw}. Additionally, such late decay of DM into photons can cause energy injection into the plasma during the post-recombination era causing spectral distortions in the cosmic microwave background (CMB). While the latest data from PLANCK mission can constrain such CMB spectral distortions, future CMB experiments will be able to probe them further by several order of magnitudes \cite{Liu:2023nct}.

Another detection prospects via CMB measurement is the presence of excess dark radiation in our setup. The requirement of keeping Dirac Yukawa interaction of neutrinos in equilibrium prior to the sphaleron decoupling epoch also thermalises $\nu_R$ enhancing the effective relativistic degrees of freedom or $N_{\rm eff}$. The current bound on such excess dark radiation or $\Delta N_{\rm eff} = N_{\rm eff}-N^{\rm SM}_{\rm eff}$ from PLANCK 2018 measurement reads $\Delta N_{\rm eff} < 0.285$ at $2\sigma$ \cite{Planck:2018vyg} where $N_{\rm eff}^{\rm SM}=3.045$\cite{Mangano:2005cc, Grohs:2015tfy,deSalas:2016ztq}. Future CMB experiments like CMB-S4 and CMB-HD are sensitive to $\Delta N_{\rm eff} = 0.06$ \cite{Abazajian:2019eic} and $\Delta N_{\rm eff} = 0.014$ \cite{CMB-HD:2022bsz} respectively. Thermalised light Dirac neutrinos with decoupling temperature above electroweak scale leads to $\Delta N_{\rm eff} =0.14$ \cite{Abazajian:2019oqj}, assuming three copies of $\nu_R$. Therefore, our Dirac cogenesis scenario remains within reach of multiple CMB missions due to enhanced $\Delta N_{\rm eff}$.

The heavy neutral lepton (HNL) like $N$ around the TeV corner can be produced at colliders via dipole interactions. If sufficiently produced, such HNL can give rise to mono-photon signatures by decaying within the detector. A recent review of such HNL searches can be found in \cite{Abdullahi:2022jlv}. In addition to these possibilities of probing our cogenesis setup, it can also be falsified by future observation of neutrinoless double beta decay which rules out pure Dirac nature of light neutrinos.

\section{Conclusion}
\label{sec4}
We have proposed a novel cogenesis mechanism where asymmetries in lepton and dark sectors are generated from decay of heavy vector-like fermions facilitated by electromagnetic dipole interactions. The absence of any net lepton number violation not only predicts the light neutrinos to be of Dirac type, but also fixes the mass of DM to be approximately twice as heavy as the proton. Unlike canonical leptogenesis in seesaw models of light neutrino masses, here it is possible to have successful cogenesis at a scale as low as a TeV without requiring resonant enhancement. Keeping washouts out-of-equilibrium puts strong constraints on the parameter space of the model pushing the cutoff scale $\Lambda$ in dimension-5 dipole operators to higher values, heavier than the scale of leptogenesis by several order of magnitudes. While the model can be falsified by future observation of neutrinoless double beta decay, three-body decay of DM into neutrino and two-photon final states can be probed at future gamma-ray telescopes. Future CMB missions can also probe parts of the model parameter space causing spectral distortion from DM decay. On the other hand, thermalised Dirac neutrinos lead to enhanced dark radiation $\Delta N_{\rm eff} \geq 0.14$, within reach of several proposed CMB experiments. In addition to these tantalising prospects at experimental frontiers, our proposed setup should also offer model building avenues which we skip here for the sake of minimality by adopting an effective field theory setup. We leave such details to future works.

\acknowledgements
The work of D.B. is supported by the Science and Engineering Research Board (SERB), Government of India grants MTR/2022/000575 and CRG/2022/000603. D.B. also acknowledges the support from the Fulbright-Nehru Academic and Professional Excellence Award 2024-25.


\providecommand{\href}[2]{#2}\begingroup\raggedright\endgroup

\end{document}